\documentclass[twocolumn, showpacs, showkeys]{revtex4}

\usepackage{graphicx}
\usepackage{graphics}
\usepackage{amsmath}
\usepackage{amssymb}
\usepackage{float}

\begin{document}
\title{Study of the island morphology at the early stages of Fe/Mo(110) MBE growth}

\author{Dmytro Goykolov}
\email{goykolov@fzu.cz}
\affiliation{Institute of Physics, Academy of Science of the Czech Republic, Na Slovance 2, Prague, 18221, Czech Republic}
\author{Miroslav Kotrla}
\affiliation{Institute of Physics, Academy of Science of the Czech Republic, Na Slovance 2, Prague, 18221, Czech Republic}
\begin{abstract}

We present theoretical study of morphology of Fe islands grown at Mo(110) surface in sub-monolayer MBE mode. We utilize atomistic SOS model with bond counting, and interactions of Fe adatom up to third nearest neighbors. We performed KMC simulations for different values of adatom interactions and varying temperatures. We have found that, while for the low temperature islands are fat fractals, for the temperature $500K$ islands have faceted rhombic-like shape. For the higher temperature, islands acquire a rounded shape. In order to evaluated qualitatively morphological changes, we measured averaged aspect ration of islands. We calculated dependence of the average aspect ratio on the temperature, and on the strength of  interactions of an adatom with neighbors.

\end{abstract}


\maketitle

\section{Introduction}
\label{intro}

Iron films grown on molybdenum are known to have interesting ferromagnetic properties \cite{fruchart-jubert-review}. 
Hence, nature and formation of the iron islands at the early stages of molecular beam epitaxy (MBE) growth are attracting great amount of researchers' attention. Knowing the processes of island formation in sub-monolayer regime, will provide film manufacturers the ability to control the growth process, predict the outcome of deposition and produce films with needed properties.

A number of experimental investigations were done on this subject. For the review of experimental works on the morphology and magnetic behavior of the iron islands we address the reader to \cite{fruchart-jubert-review, aballe2007} and references therein. 

Theoretical aspect of the early stages of MBE growth on bcc(110) surface still remains an open question. Some work has been done to develop a model that reproduces some experimental results for homoepitaxy on Fe(110) \cite{schindler}. Recently, results of the phase-field simulation of stripe arrays growth on bcc(110) surface in sub-monolayer regime has been reported \cite{yu-stripe}. But, to our knowledge, there are no deep theoretical investigation of the problem of island formation for Fe/Mo(110) system. 

The main goal of this work is to study morphology of the islands in the sub-monolayer regime. Here, we present some preliminary results on the island formation at early stages of Fe/Mo(110) MBE growth. The structure of the paper is as follows. In Section~\ref{model}, we give short theoretical description of MBE growth model. To implement the model we are using on-lattice Kinetic Monte Carlo (KMC) algorithm, which is widely used for modelling of MBE growth. 
In the beginning of Section~\ref{aar_vs_t}, we define quantities utilized for characterization of island morphology. Then this section continues by description of changes of island morphology while varying substrate temperature. Quantitative parameter (average aspect ratio) as well as concept of island compactness are used. In Section~\ref{aar_vs_e}, we study the morphology of the island as a function of different values of interaction energies.
We present results for the dependence of island aspect ratio on interaction energies with nearest and second nearest neighbors. Each section also contains visual illustrations of islands shape for different parameters. Section~\ref{conclusions} concludes our current results and outlines the future directions of work. 

\section{Model}
\subsection{Growth Model}
\label{model}

In our study, we used growth model based on solid-on-solid (SOS) model of molecular beam epitaxy (MBE). In original SOS \cite{weeks-sos}, the system is represented as a simple cubic crystal. No vacancies and overhang are allowed. In this case, surface may be described as two-dimensional matrix. Indexes of the matrix serve as spatial coordinates of the substrate. Value of the matrix element is the height
- number of atoms above the substrate in column at this particular coordinate.
Usually, three kind of processes are considered during the growth: deposition, surface diffusion and evaporation. We do not allow atoms to evaporate in our model. 

Deposition of atoms occurs at a random location on the substrate with the rate
\begin{equation}
\label{eq:dep-rate}
 \tau = \frac{1}{Fa^{2}},
\end{equation}
where $F$ is the incoming beam flux of deposited material and $a$ is the distance between two nearest neighbors on the lattice.

Surface diffusion is described by the set of hopping rates given by the Arrhenius distribution:
\begin{equation}
\label{eq:dif-rate}
 r_i = \nu_0e^{-E_i/k_BT},
\end{equation}
where $r_i$ is a hopping rate of adatom in configuration $i$, $\nu_0$ - attempt frequency, $k_B$ is the Boltzmann's constant, $T$ - temperature of the substrate. Energy barrier $E_i$ is calculated by the simple bond-counting scheme with the generic form:
\begin{equation}
\label{eq:energy}
 E_i = E_s + \sum_{j = 1}^{J} n_j^{(i)}E_{j},
\end{equation}
where $E_s$ is the interaction energy of the free adatom with the substrate material. 
$J$ is the total number of neighbors considered. $n_j^{(i)} \in (0,1)$, $j = 1,...,J$ is the number of bonds to the considered neighbor (which may be 0, when there is no neighbor, and 1 otherwise) in a configuration $i$. $E_{j}$ are corresponding binding energies.

In our model, we consider interactions between the adatom and its lateral first, second and third nearest neighbors. In the case of bcc(110) surface, there are 4 first nearest, 2 second nearest, and 2 third nearest neighbors (see Figure \ref{fig:1}). Hence, $J = 8$. Interaction energies $E_{j}$ are material-related parameters of the model. Further in the text we will use notations $E_n,\; E_{2n},\; E_{3n}$ for interaction energies with first, second and third nearest neighbors correspondingly. Energies for the $Fe$ on $Mo$ system were calculated 
by Yang and Asta \cite{asta} employing spin-polarized electronic density functional theory, generalized-gradient approximation \cite{perdew} calculations. Details of these calculations will be discussed in a future publication. Other model parameters do not depend on the material: substrate temperature, incoming flux of the deposited material and the surface coverage.


In more detailed models, such effects as possibility of edge diffusion of an adatom, influence of a step-edge on interlayer transport, or presence of the strain (due to lattice mismatch between substrate and deposited material), should be   taken into account. However, these effects are not consider at this stage of our work.


\subsection{Algorithm}
\label{algorithm}

The most common way to simulate MBE growth is to use Kinetic Monte Carlo algorithm, also referenced to as BKL. It was proposed by Bortz \textit{et al.} in 1975 \cite{bkl}. In the core of KMC algorithms lie two key mechanisms. The first is the keeping of the list of all possible events in the system. The second one is the selecting particular event by making a search through the list of events. Since the appearance of BKL paper several different versions of KMC algorithm have been developed, for review see e.g. \cite{kotrla96}.
Different versions use different approach to organizing event lists and different search algorithms. The most used KMC algorithms are the algorithm with linear search (LS), binary search (BS) (which is useful in case of large number of events and differs from LS algorithm by the method of bookkeeping: the list of events is organized in form of the binary tree), K-level search \cite{kls} and, finally, binning method by Maksym \cite{maksym}.

For our simulations we adopt Maksym's method modified by Haider \textit{et al.} \cite{haider} which allows to prevent events with high hopping rates from dominating the simulation. In the following, we give short description of the algorithm with implemented specifications to suit our model.


\begin{figure}
 \centering
 \includegraphics[width=0.25\textwidth,bb=127 344 409 713]{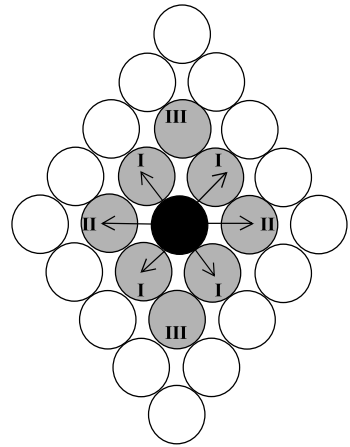}
 \caption{Fragment of the bcc(110) surface with allowed directions of jumps of the adatom (dark circle). Lateral neighbors that interact with the adatom (nearest, second nearest and third nearest neighbors) are shown as light-gray circles and are marked by corresponding roman numerals. Arrows show allowed directions of the jump.}
 \label{fig:1}
\end{figure}


Fragment of bcc(110) surface is shown on Figure \ref{fig:1}. As was mentioned before, the surface is represented as two-dimensional array with indices being the coordinates of the substrate site and the value of the array element is the number of deposited adatoms at that location. To eliminate any preferences in selecting particular sites we create a structure, that contains pointers to the random lattice sites. In order to speed up the search procedure, lattice is divided into a number of smaller groups. Each group contains pointer to the first element of the site list  in this group, and number of events of a certain type. And again, to avoid any preferences in selecting certain group, we create a structure that contains pointers to random groups.

In order to select particular event and location of this event, first we need to calculate total rate of all events in the system. We calculate this rate as a sum of all possible event rates weighted with the number of atoms in each local configuration type:
\begin{equation}
\label{eq:total-rate}
 R(N_1,...,N_M) = N\tau + \sum_{\alpha=1}^M N_\alpha r_\alpha,
\end{equation}
where $N$ is the total number of surface atoms, $\tau$ is the deposition rate \eqref{eq:dep-rate}, $M$ is the total number of configuration types, $N_\alpha$ - number of atoms with configuration type $\alpha$, $r_\alpha$ - hopping rate of the configuration $\alpha$ \eqref{eq:dif-rate}. 
In our algorithm, individual hopping rates are calculated using the following diffusion barrier of the adatom:
\begin{equation}
 \label{eq:dif-barrier}
 E = E_s + n_1E_n + n_2E_{2n} + n_3E_{3n},
\end{equation}
where $n_1,\; n_2,\; n_3$ are the counts of number of occupied first, second and third nearest neighbors correspondingly. Taking into account geometry of the lattice (Figure~\ref{fig:1}), there are 5 combinations of the nearest neighbors configurations, in which number of neighbors can be from 0 to 4, and 3 combinations of second and third nearest neighbors, where number of neighbors can be from 0 to 2. Hence, total number of different configuration types $M=45$.


Individual probabilities of events may be calculated as follows:
\begin{equation}
\label{eq:prob}
\begin{gathered}
 P(\tau) = N\tau/R,\\\
 P(r_\alpha) = N_\alpha r_\alpha/R,
 \end{gathered}
\end{equation}
where $P(\tau)$ is the probability of the deposition event and $P(r_\alpha)$ is the probability of the hopping event with configuration $\alpha$. Then a cumulative probability table is defined:
\begin{equation}
\label{eq:cumulaive-prob}
 \begin{gathered}
  C_\alpha = \sum_{j=1}^\alpha P(r_j),\; \alpha=1,...,M \\
  C_{M+1} = C_M + P(\tau).
 \end{gathered}
\end{equation}
In order to select an event a random number $\Re_1$ in the region $[0,R)$ is generated. Event $\alpha$ is then selected for which

\begin{equation}
\begin{gathered}
C_{\alpha-1} < \Re_1 \leqslant C_\alpha, \quad \mbox{for hopping event,} \\
C_M < \Re_1 \leqslant C_{M+1}. \quad \mbox{for deposition event.}
\end{gathered}
\end{equation}
From this equation we can see, that if $\Re_1 > C_M$ then deposition event will be realized. In other case one of the hopping events will be selected.

The last step is to select the lattice site at which chosen event occurs. In case of the hopping event it is done in two substeps. First - search through the list of groups to find a group, that contains non-zero amount of events of a given type. To suppress any preferences in selecting groups, every time the search starts from the random group. After the group is found, the search continues inside the group (also, every time it starts from the random element of the group). 
The destination site for the jump is chosen randomly out of 6 adjacent neighbors. 
In our algorithm, we allow adatom to jump to its lateral first and second nearest neighbors (see Figure \ref{fig:1}). This choice was made mainly because of the geometrical considerations: first and second nearest neighbors are almost at the same distance from the jumping adatom. Besides, allowing atoms to jump to the third nearest neighbors sites would considerably increase the complexity of the algorithm realization.
In case of the deposition event, site is chosen randomly.

After the performing an event, it is needed to update the simulation timer. Since events are determined by a uniformly distributed random numbers, time interval between two consecutive events $t$ is derived from the Poisson distribution of time intervals:
\begin{equation}
 P(t)dt = Re^{-Rt}dt,
\end{equation}
where $R$ is the total rate of all events. From this distribution we can derive the time interval between events \cite{voter}:
\begin{equation}
 t = -\frac{\log\Re_2}{R},
\end{equation}
where $\Re_2$ is a random number between 0 and 1, $R$ - total rate of all possible events.

Simulation continues until the terminal value of surface coverage is reached. Particular conditions of each simulation will be given during their discussion.

\section{Results}
\subsection{Dependence of island morphology on the substrate temperature}
\label{aar_vs_t}

In order to evaluate change of island shape, we need to define some quantitative parameter, therefore, we introduce average aspect ratio (a.a.r.). It is calculated as an average of ratio of transversal and longitudinal dimensions for each island averaged over all atomic islands in the system. Another quantity, that we will use in our discussion is the island compactness. It will be determined visually - if island has a shape of a fractal then it is not compact. 
A fractal is a rough or fragmented geometric shape that can be subdivided in parts, each of which is, in a statistical sense, a reduced-size copy of the whole. Fractals are generally self-similar and independent of scale. We will refer to the fractals that are less fragmented as to fat fractals. We will call island compact if it is not a fractal.


\begin{figure}[!t]
 \centering
 \includegraphics[width=0.45\textwidth]{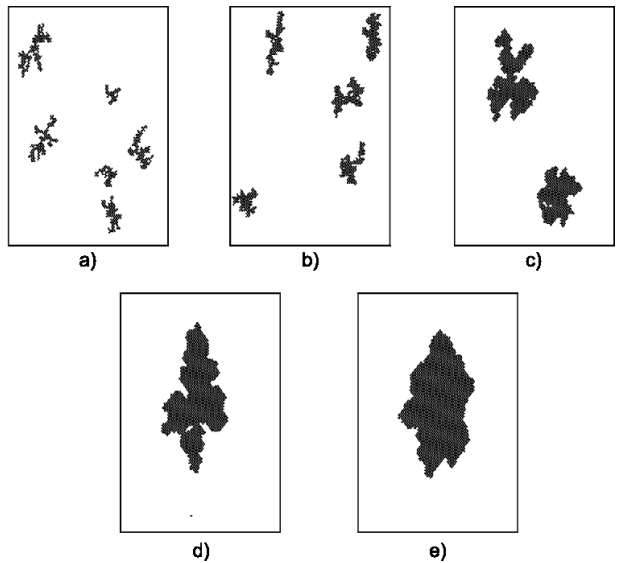}
 \caption{Change of island shape with temperature growth: a) $T=325K$, b) $T=350K$, c) $T=400K$, d) $T=425K$, e) $T=450K$. Conditions of simulations are given in the text.}
 \label{fig:shape_t}
\end{figure}

Figure \ref{fig:shape_t} shows examples of the islands that were grown for different temperature of the substrate. Except for temperature, all other simulation parameters were the same: lattice size $300\times300$, incoming beam flux $F = 0.01\; ML/s$ (monolayers per second; this value is constant for all simulations), final surface coverage $\theta = 0.05\; ML$, atomic vibrational frequency $F_{vib} = 4\times10^{12}\; Hz$, and interaction energies are $E_s=0.4\; eV,\; E_n = 0.329\; eV,\; E_{2n}=0.072\; eV,\; E_{3n}=0.079\; eV$.

Hopping rates of the atoms increase with the temperature growth. Consequently, overall time spent for the surface diffusion also is growing with the temperature. That means that during the deposition on the substrate with the higher temperature atoms will have more time to diffuse to energetically preferable locations. Those locations are the ones with the maximum number of lateral neighbors. Hence, for higher temperatures islands are becoming more compact. 

\begin{figure}[!t]
 \centering
 \includegraphics[width=0.45\textwidth]{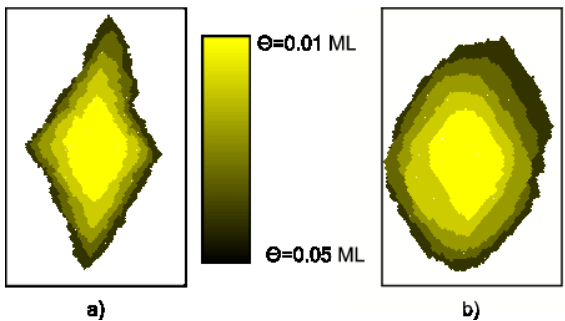}
 \caption{Color online. Island shapes and evolution with coverage for two temperature values. a) $T=500K$, calculated energies; b) $T=350K$, scaled energies (see explanation in the text). Fragment of the lattice was scaled down.}
 \label{fig:shape_t_color}
\end{figure}


Continuation of the shape evolution is shown on the Figure \ref{fig:shape_t_color}. First island snapshot is the island at $T=500K$ and the same conditions described above. To eliminate finite size effect, lattice size was increased to $1000\times1000$. We can see that an island is clearly bordered by two kinds of facets. Figure \ref{fig:shape_t_color}b shows the island obtained under another conditions. For this simulation all energies and the substrate temperature were scaled by a factor $\alpha = 0.5$. Initially, scaling was done in order to reduce computation time. Multiplying both energies and temperature by the same factor does not change the hopping rates of the atoms \eqref{eq:dif-rate}. Hence, we may say that the same result would be obtained for initial set of energies (Figures \ref{fig:shape_t} and \ref{fig:shape_t_color}a) and temperature $T=700K$. Figure~\ref{fig:shape_t_color} was scaled to fit the picture. Originally, the size of this rounded island is much larger that the size of rhombic-like island on the Figure~\ref{fig:shape_t_color}a. 

One can observe three stages of the shape change on the Figures \ref{fig:shape_t} and \ref{fig:shape_t_color}: from $T=325K$ to $T=400K$ islands are changing from being fractals to fat fractals. After that the islands shape is becoming compact quadrangle at $T=500K$. As temperature growth further, islands start to loose their distinc facets and obtain rounded shape as shown on the Figure \ref{fig:shape_t_color}b. Similar islands with rounded shapes were observed in the experimental work \cite{bode-rt, prokop-rounded}.

Another type of shape evolution of the islands may be observed on the Figure \ref{fig:shape_t_color}: evolution of the island shape with the coverage. We can observe, that islands obtain their shape at the very early stages of the growth. As the deposition proceeds, islands only grow in size and do not change the shape. Of course, there is a possibility for two or more islands to merge. In this case (shortly after the merge) the resulting island will not have the shape of other individual islands. But given enough time for the atoms to diffuse, merged island shape will evolve into rhombic-like or rounded island. This is not the case for the fractal and fat fractal islands. Since atoms are less mobile on the substrate with low temperatures, fractal islands will remain fractals after the merge.

 \begin{figure}[!t]
  \centering
  \includegraphics[width=0.45\textwidth]{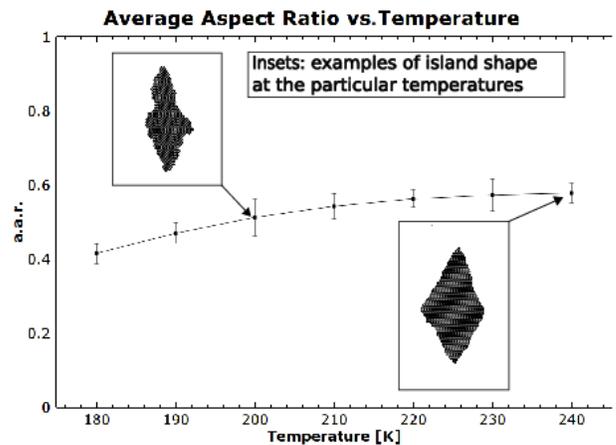}
  \caption{Dependence of the average aspect ratio on the substrate temperature. Insets show examples of island shapes at different temperatures. Conditions of the simulations are given in the text.}
  \label{fig:aar_vs_t}
 \end{figure}

Figure \ref{fig:aar_vs_t} shows dependence of the a.a.r. on the substrate temperature. Data for this graph was obtained under the following simulation conditions: for temperatures $T=180..200K$ lattice size was $300\times300$, for $T=210..230K$ - $600\times600$ and for $T=240K$ size was $1000\times1000$, final surface coverage $\theta = 0.1\; ML$; atomic vibrational frequency $F_{vib} = 2\times10^{12}\; Hz$; interaction energies $E_s = 0.2\; eV,\; E_n = 0.14\; eV,\; E_{2n} = 0,\; E_{3n} = 0.06\; eV$. For $300\times300$ lattices data was averaged over 10 independent simulations and for larger lattices - over 5 independent runs.

Dependence of average aspect ration on the temperature that is shown on the Figure \ref{fig:aar_vs_t} supports our previous observations on the island shape. As temperature growth, islands' average aspect ratio is growing, indicating that islands becoming more compact. This statement is also illustrated by the insets on the Figure~\ref{fig:aar_vs_t}. These insets show examples of the islands at different temperatures and we can see the transition from the islands without sharp facets to the compact rhombic-like islands.

\subsection{Dependence of island morphology on the interaction energies}
\label{aar_vs_e}

Temperature is not the only parameter, that has an influence on the island shape. In this section we present the results on how the interaction energies of the adatom with its lateral neighbors change the islands shape. Since in the following simulations we are changing material-dependent parameters (interaction energies), this section is not strictly related to the Fe/Mo(110) growth process. Nevertheless, these results may give usefull insight to the island morphology for the systems with other sets of interaction energies.

\begin{figure}[!h]
 \centering
 \includegraphics[width=0.45\textwidth]{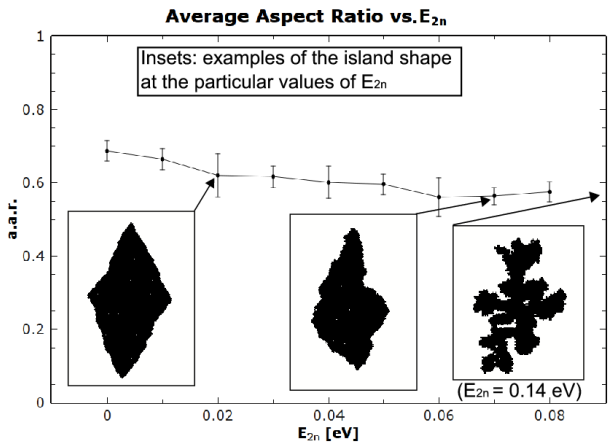}
 \caption{Dependence of average aspect ratio on interaction energy with second nearest neighbor. Insets show islands at different values of $E_{2n}$}
 \label{fig:aar_vs_E2n}
\end{figure}

Figure \ref{fig:aar_vs_E2n} shows the dependence of average aspect ration on the interaction energy of the adatom with the second neareast neighbors. To obtain this result, the following simulation conditions were used: $300\times300$ lattice, $\theta = 0.1\; ML$, $T = 230K$, $F_{vib} = 2\times10^{12}\; Hz$, interaction energies: $E_s = 0.2\; eV,\; E_n = 0.14\; eV, \; E_{3n} = 0.03\; eV$. The data was averaged over 10 independent runs.

As one can see, the main influence of $E_{2n}$ is not the change of a.a.r., but the change of shape of the islands. The shape is changing from compact at low values of $E_{2n}$ to the shape without sharp facets and, eventually, to the fractals. This evolution is illustrated by the insets on the Figure \ref{fig:aar_vs_E2n} which are showing snapshots of the islands at different values of $E_{2n}$. Simulations were terminated at $E_{2n}=0.08\; eV$ since around that value island shape started to become fat fractal and measuring a.a.r. is usefull only for compact islands.

\begin{figure}[!h]
 \centering
 \includegraphics[width=0.45\textwidth]{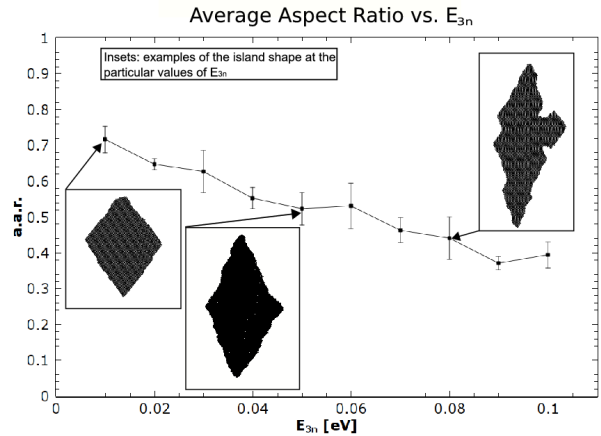}
 \caption{Average aspect ratio as a function of interaction energy with third nearest neighbor. Insets show islands at different values of $E_{3n}$}
 \label{fig:aar_vs_E3n}
\end{figure}

Dependency of the island morphology on $E_{3n}$ is shown on the Figure \ref{fig:aar_vs_E3n}. Here we used $300\times300$ lattice, terminal coverage $\theta=0.1\; ML$, substrate temperature $T=250K$, atomic vibrational frequency $F_{vib} = 2\times10^{12}\; Hz$ and interaction energies $E_s = 0.2\; eV,\; E_n=0.14\; eV,\; E_{2n}=0.04\; eV$. Each data point was averaged over 10 independent runs.

Influence of $E_{3n}$ on the island shape appears to be stronger than that of $E_{2n}$. Although one can observe the decrease of a.a.r. with growth of $E_{3n}$ (as in the case of changing $E_{2n}$), the change of island shape differs from the case above. Here with the change of $E_{3n}$ islands become more prolongated which leads to the decrease of average aspect ratio. But at the same time the compactness of the islands is preserved - they are not becoming fractals or even fat fractals in the range of simulations. Change of the island shape is also illustrated by the insets on the Figure \ref{fig:aar_vs_E3n}, where one can see islands at different values of $E_{3n}$.

\section{Conclusions}
\label{conclusions}

In this paper, we have presented results of KMC simulations of sub-monolayer island growth at bcc(110) surface. We have utilized extended bond counting model with interactions up to third nearest neighbors. Interactions were based on \textit{ab initio} calculations. However, for this parameter set we were able to perform simulations only for temperatures up to $500K$. In order to obtain results for higher temperatures we used scaled interaction energies. 

We observed usual transition of island shape from fractal to compact. Compact islands at $T=500K$ have rhombic-like shape. For higher temperatures they obtain rounded shape in agreement with the experiment \cite{bode-rt, prokop-rounded}. In order to quantify morphological change we have evaluated the dependence of average aspect ratio of island on the temperature and values of adatom interactions. We observed that with increasing temperature aspect ratio is increasing, and eventually saturates. In the case of dependence on the strength of interaction, we found that the aspect ratio decreases slowly with the increase of interaction energy with the second neighbor. But the island shape is quickly transforming from compact to fat fractals. With the increase of interaction energy with the third neighbor, the aspect ratio is decreasing much faster than in the previous case. At the same time, the island shape remains compact in the range of simulation.

In addition, we observed that the system may have a state where the islands develop an extra facet and resemble hexagon-like shapes. This our result is similar to the experimental works where hexagon islands were observed \cite{jubert-hexagons}. However, the conditions (energies, temperatures and coverage) under which islands with hexagon-like shapes are grown are still needed to be clarified. In the case of heteroepitaxy, it is also important to consider effect of strain which is currently not included in our model. In order to obtain more realistic results we are working on the new algorithm which will include edge diffusion and strain.

We have realized that our calculations for high temperatures ($T>500K$) and interaction energies obtained by \textit{ab initio} calculation are very demanding in terms of computer time. Nevertheless, our results are in qualitative agreement with experimental findings.

\textbf{Acknowledgments:} The work was supported by joint funding under EU STRP 016447 MagDot and NSF DMR Award No. 0502737, and by Project AVOZ 10100520 of ASCR. We thank M. Asta for providing calculated interaction energies.


\begin{thebibliography}{5}
 \bibitem{fruchart-jubert-review} O. Fruchart et al., J.Phys.:Condens. Matter 19, 053001 (2007)
 \bibitem{aballe2007} L. Aballe, A. Barinov, A. Locatelli, T. O. Montes, and M. Kiskinova, Phys. Rev. B 75, 115411 (2007)
 \bibitem{schindler} U. K\"{o}hler, C. Jensen, A. C. Schindler, L. Brendel, and D. E. Wolf, Phil. Mag. B 80, 283 (2000)
 \bibitem{yu-stripe} Y.-M. Yu, Rainer Backofen and Axel Voight, Phys. Rev. E 77, 051605 (2008)
 \bibitem{weeks-sos} J. D. Weeks and G. H. Gilmer, Adv. Chem. Phys. 40, 157 (1979)
 \bibitem{asta} M. Asta, private communacation.
 \bibitem{perdew} J. P. Perdew, K. Burke and M. Ernzerhof, Phys. Rev. Lett. 77, 3865 (1996).
 \bibitem{bkl} A. B. Bortz, M. H. Kalos and J. L. Lebowitz, J. Comp. Phys. 17, 10 (1975)
 \bibitem{kotrla96}M. Kotrla, Comp. Phys. Comm. 97, 82 (1996)
 \bibitem{kls}J. L. Blue, I. Beichl and F. Sullivan, Phys. Rev. E 51, 867 (1995)
 \bibitem{maksym} P. A. Maksym, Semicond. Sci. Technol. 3, 594 (1988)
 \bibitem{haider}N. Haider, S. A. Khaddaj, M. R. Wilby and D. D. Vvedensky, Comput. Phys. 9, 85 (1995)
 \bibitem{voter}A. F. Voter, Introduction to the Kinetic Monte Carlo Method, in Radiation Effects in Solids, edited by K. E. Sickafus and E. A. Kotomin (Springer, NATO Publishing Unit, Dordrecht, The Netherlands, 2005) 
 \bibitem{bode-rt} M. Bode, O. Pietzsch, A. Kubetzka, and R. Wiesendanger, Phys. Rev. Lett. 92, 067201 (2004)
 \bibitem{prokop-rounded} J. Prokop, A. Kukunin, and H. J. Elmers, Phys. Rev. B 73, 014428 (2006)
 \bibitem{jubert-hexagons} P. O. Jubert, O. Fruchart, and C. Meyer, Phys. Rev. B 64, 115419 (2001)
\end{thebibliography}
\end{document}